# Tunable lasers with optical-parametric oscillation in photonic-crystal resonators

**JENNIFER A. BLACK,**[1,*] **GRANT BRODNIK,**[1,2] **HAIXIN LIU,**[1,2] **SU-PENG YU,**[2] **DAVID R. CARLSON,**[1,3] **JIZHAO ZANG,** [1,2] **TRAVIS C. BRILES,**[1] **AND SCOTT B. PAPP**[1,2]

[1] *Time and Frequency Division, National Institute of Standards and Technology, Boulder, Colorado 80305, USA*
[2] *Department of Physics, University of Colorado, Boulder, CO 80309, USA*
[3] *Octave Photonics, Louisville, CO 80027*
*Corresponding author: jennifer.black@nist.gov*

**By design access to laser wavelength, especially with integrated photonics, is critical to advance quantum sensors like optical clocks and quantum-information systems, and open opportunities in optical communication. Semiconductor-laser gain provides exemplary efficiency and integration but merely in developed wavelength bands. Alternatively, nonlinear optics requires control of phase matching, but the principle of nonlinear conversion of a pump laser to a designed wavelength is extensible. We report on laser-wavelength access by versatile customization of optical-parametric oscillation (OPO) with a photonic-crystal resonator (PhCR). By controlling the bandgap of a PhCR, we enable OPO generation across a wavelength range of 1234–2093 nm with a 1550 nm pump and 1016–1110 nm with a 1064 nm pump. Moreover, our tunable laser platform offers pump-to-sideband conversion efficiency of >10% and negligible additive optical-frequency noise across the output range. From laser design to simulation of nonlinear dynamics, we use a Lugiato-Lefever framework that predicts the system characteristics, including bi-directional OPO generation in the PhCR and conversion efficiency in agreement with our observations. Our experiments introduce tunable lasers by design with PhCR OPOs, providing critical functionalities in integrated photonics.**

http://dx.doi.org/10.1364/OL.99.099999

Advances in laser applications, especially in fluctuating ambient conditions and with less developed spectral ranges, drives innovation in laser and integrated photonics technologies. For example, quantum-based sensors [1], quantum-information systems [2], atomic and molecular spectroscopy [3], data communication [4], and photonic-signal generation [5] require diverse laser-wavelength access to enhance performance and enable novel application opportunities. Hence, the development of widely deployable laser sources that support wavelength customization would be broadly utilized. Laser technologies based on solid-state materials, doped optical fibers, and semiconductor gain with bulk cavities are commercially mature with well-understood cost and performance trade-offs. Monolithically integrated semiconductor lasers satisfy a substantial range of applications; however, no semiconductor material offers sufficiently low optical loss to support narrow spectral linewidth commensurate with bulk and fiber lasers [6]. Heterogeneously integrated lasers with fabrication on a common substrate have been developed in widely used spectral ranges and leveraging low loss waveguide materials has enabled narrow linewidth and ultraprecise laser stabilization [7]. However, heterogeneous laser fabrication is exceptionally complex, limiting wavelength access. In the context of these developments, nonlinear wavelength conversion of a pump source by use of integrated photonics is a flexible tool to expand the palette of any laser platform.

Optical-parametric oscillation (OPO) in Kerr microresonators is ubiquitous, transforming the flat state of a continuous-wave pump laser to the Turing pattern composed of a few wavelengths [8–10]. In this case, phase-matching for OPO is intrinsic, owing to the balance between Kerr frequency shifts and anomalous group-velocity dispersion (GVD). Moreover, the Turing pattern is merely one portion of the phase diagram that describes extended patterns and states of the intraresonator field that may be leveraged for tunable laser and frequency-comb generation [11–13]. Through GVD engineering via waveguide geometry in advanced integrated nonlinear photonics platforms, it is possible to control the phase-matching that determines the output frequency of a tunable OPO laser. This technique in Kerr microresonator OPOs have been explored with silica [14,15], aluminum nitride [16], and silicon nitride [17,18]. OPOs with integrated $\chi^{(2)}$ microresonators have also been explored through a more challenging phase-matching process of the pump, signal and idler [19].

Here we introduce a tunable-laser platform based on degenerate four-wave-mixing OPO in a PhCR. Our devices are inscribed with an edgeless nanopattern on the inner wall that opens a photonic bandgap (BG) for a single azimuthal mode. By coupling forward and backward propagation of one azimuthal mode, we obtain two modes with a frequency shift to higher and lower frequency compared with the unperturbed mode, respectively. Either the higher or the lower frequency split mode directly enables OPO phase matching with respect to a pair of other azimuthal modes to generate signal and idler waves. Hence, the PhCR BG is a single parameter that we use to provide laser-wavelength access across >100 THz from a single-frequency pump laser. Moreover, the forward and backward coupling associated with the PhCR BG creates bi-directional propagation of the OPO tunable laser that we explore with experiment and theory. We report on the properties of this tunable laser platform, including high conversion efficiency of >10%, low additive frequency noise in wavelength conversion, and precision output frequency tuning.

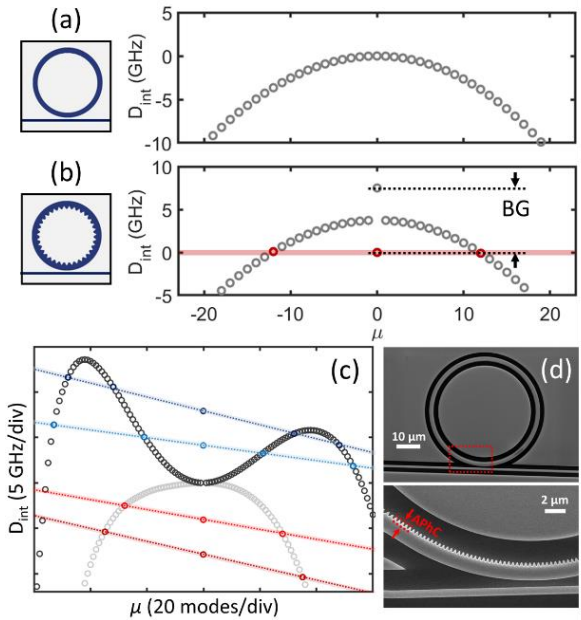

Fig. 1. (a) Conventional microresonator schematic and normal GVD $D_{int}$ which does not support OPO phase-matching. (b) PhCR schematic and $D_{int}$ where a BG is open at the pump, enabling phase-matching for OPO on the red-shifted PhCR mode. The dotted line highlights the OPO modes with optimal phase-matching. (c) Two example resonator geometries with $D_{int}$ including higher-order dispersion terms. Two settings of BG are shown with OPO phase-matching (dotted lines) for the blue- and red-shifted PhCR modes. (d) Scanning electron microscope image of a full PhCR and zoom-in of the PhCR showing the modulation amplitude, APhC.

Phase-matching for OPO is intrinsic in a microresonator when the signal and idler azimuthal mode numbers are symmetric with respect to the pump-laser mode. Frequency-matching of resonator modes therefore naturally dictates the OPO output frequencies, and GVD and Kerr frequency shifts are the principal contributions [9].

We quantify OPO frequency matching with the integrated dispersion,

$$(1) \quad D_{int}(\mu) = \nu_\mu - (\nu_0 + FSR \cdot \mu),$$

where $\nu_\mu$ are the resonator mode frequencies, $\mu$ is the mode number relative to the pump laser ($\mu = 0$), and FSR is the free-spectral range. We don't account for the Kerr frequency shift, which is twice as large for $\mu \neq 0$ than the pump mode [10]. The anomalous GVD regime is specialized for OPO frequency matching, since $D_{int}$ is positive near the pump, whereas normal GVD doesn't assist in frequency matching. As we show below, the purpose of our PhCR is to open a BG that uniquely phase matches OPO under a variety of GVDs, providing a mechanism to design tunable lasers for wavelength access. For example, the engineered BG of an otherwise normal GVD PhCR directly phase matches OPO.

Degenerate four-wave-mixing based OPO conserves energy and requires phase-matching to support appreciable parametric gain. Resonator modes with perfect phase- and frequency-matching satisfy the equation:

$$(2) \quad D_{int}(\mu) = -D_{int}(-\mu).$$

Practical device design seeks to minimize frequency-mismatch of modes symmetric about the pump to less than a resonator linewidth ($\delta$), $D_{int}(\mu) + D_{int}(-\mu) < \delta$. We focus on explaining microresonator OPO engineering with normal GVD in Fig. 1a. The quantity $D_{int}$ is zero by construction at the pump mode ($\mu = 0$), and we find that $D_{int}(\nu_{\mu\neq 0}) < 0$, which can't satisfy equation 2. However, an analogous PhCR opens a bandgap on a single resonator mode; see Fig. 1b. Then by pumping the red-shifted (ie. shifted to lower frequency) PhCR mode ($\mu = 0$), OPO is phase-matched at modes $\mu = +/- 12$ (shaded red box), since the pump mode is frequency shifted in a fashion consistent with an anomalous GVD regime for only the pump mode. The spacing of the generated OPO frequencies is $\delta\nu_{OPO} = \nu_s - \nu_i$, where $\nu_s$ and $\nu_i$ are the generated OPO signal and idler frequencies.

In practice, higher-order dispersion can be a significant consideration, creating the more complex $D_{int}$ trends of a realistic PhCR device; see Fig 1c. In practice, we tune the GVD using the geometry of the ring resonator. For example, the $D_{int}$ in Fig. 1c are calculated for experimental resonator geometries using tantala ($Ta_2O_5$) waveguides with thickness of 570 nm, top and side air-cladding, a $SiO_2$ bottom cladding, and a ring resonator radius (RR) of 22.5 $\mu$m, which we measure to the center of the ring waveguide. We achieve different GVD by changing the ring waveguide width (RW), which are 1625 nm and 1710 nm for black and grey $D_{int}$ traces Fig. 1c, respectively. Here, we present two settings of BG for otherwise anomalous (black) and normal (grey) GVD microresonators, and we show the OPO-matched modes, using dotted lines; red and blue lines signifying OPO on the red and blue shifted PhCR mode, respectively. The OPO-matched pump, signal and idler modes satisfy the OPO condition. As described above (Fig. 1b), we intend to pump the red-shifted PhCR mode in an otherwise normal GVD resonator, enabling OPO with increasing $\delta\nu_{OPO}$ proportional to BG. In an anomalous GVD resonator, we intend to pump the blue-shifted PhCR mode, providing OPO in two possible branches due to roll-over of $D_{int}$ at modes further from the pump.

The physical mechanism to open the BG is fabrication of a modulation on the inner wall of the microresonator. Scanning electron microscope images (Fig. 1d) show both the entire PhCR (top) and a zoomed-in image of the PhCR, highlighting the modulation of the inner wall with amplitude APhC. The magnitude of the BG depends on APhC, and the optical frequency of the PhC BG depends on the modulation period, $2 \cdot m$, where m is the azimuthal mode order of the resonator mode where the BG is open.

which use inverse tapers of length 200 μm and width 2.75 μm at the edge of the chips to best mode match to the lensed fibers. We use a straight bus waveguide geometry to couple light to the PhCR, which feature a typical intrinsic quality factor of $2 \times 10^6$.

Figure 2a presents our experimental setup for testing tunable lasers created by PhCR OPO, including assessment of bi-directional OPO generation. Our pump laser is an amplified external cavity diode laser of wavelength in either the 1064 nm or 1550 nm bands. We monitor the pump power in the forward and backward direction by using a fiber circulator before the PhCR and optical spectrum analyzers monitor the optical spectra from the PhCR in both directions. We photodetect the pump and OPO power as a function of pump laser detuning in both directions, using optical filtering. The principal characteristic of our PhCRs is BG-enabled output laser tuning in which the output is deterministically set in steps of one FSR. Figure 2b presents three sample optical spectra demonstrating tunable laser operation, where we tune the PhCR OPO output frequencies by tuning BG. The vertical dashed lines are separated by the 1 THz FSR, demonstrating BG-enabled wavelength control.

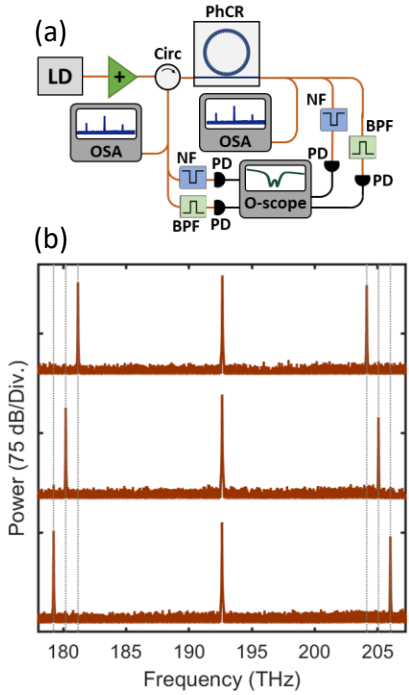

Fig. 2. (a) Experimental setup for monitoring bi-directional PhCR OPO. Power from a laser diode (LD) is amplified (+) and passed through a fiber circulator (Circ). Bi-directional OPO generation is monitored on an optical spectrum analyzer (OSA) and filtered (notch- and bandpass-filter: NF, BPF) before monitoring with photodiodes (PD) and an oscilloscope (O-scope). Orange/black paths are optical/electrical. (b) Optical spectra for three settings of BG showing FSR-scale tuning (vertical dashed lines have 1-THz FSR-spacing).

With the context of PhCR OPOs presented above, we turn our attention to fabrication and experiments that explore formation of tunable lasers and their characteristics. To create PhCRs, we use the tantala integrated nonlinear photonics platform that we have developed [10,20–22]. Tantala offers advantages for PhCR OPOs, including access to high Q in the 1064 nm, 1550 nm, and 2000 nm wavelength bands that we explore here and precision control of GVD, due to <1 nm film thickness variation across a 75 mm wafer. We use a commercial vendor, FiveNine Optics, to deposit a 570 nm thick tantala film on an oxidized silicon wafer by ion-beam sputtering, and our designs are transferred to the tantala layer through electron-beam lithography and fluorine ICP-RIE. Thermal annealing in air for several hours at 500 °C reduces oxygen vacancies present in the tantala material. Our customized fabrication process yields more than 20 chips with ~100 microresonators per chip in a focused 2-day fabrication period. We use lensed fibers to insert and collect light from the bus waveguides,

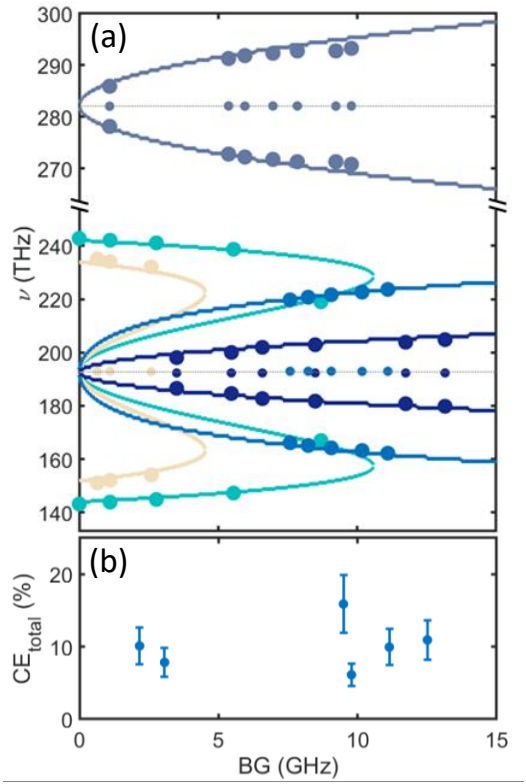

Fig. 3. (a) Summary of tunable laser output frequencies versus BG for various resonator geometries. Dots (lines) are experimental (theoretical) results. Horizontal dashed lines denote the pump frequency. (d) Total conversion efficiency versus PhCR bandgap.

Figure 3 presents the output frequency from several PhCR OPO devices as a function of BG, emphasizing laser-wavelength access with high conversion efficiency. We find that with nearly the same pump frequency (193 THz), we generate output frequencies (dots) from 143 THz to 243 THz. We also demonstrate PhCR OPO by pumping at 282 THz, accessing frequencies from 278 THz to 286

THz; see the upper broken axis in Fig. 3a. The solid lines in Fig. 3a denote the theoretical phase-matching computed with finite element method modeling and equation 2 for a few settings of resonator RW. The parabolic curves (grey, blue, dark blue) denote OPO in otherwise normal GVD resonators, and the turquoise and beige curves denote OPO in anomalous GVD resonators, demonstrating the two-branch behavior discussed above. Since the PhCR OPO is bi-directional, we observe laser generation in both the forward and backward direction, hence we monitor the power propagating in both directions to determine the conversion efficiency from the pump to the OPO output. We calculate the total conversion efficiency ($CE_{total}$) as the sum of the optical power generated in the forward and backward direction compared to the input optical power. Note that at higher pump powers, cascaded OPO and other saturation effects that trend toward a frequency comb are possible [23], therefore in these measurements we only consider optical power at our target OPO frequencies. We plot $CE_{total}$ in Fig. 3b as a function of BG for resonators pumped to ~1.3 times the OPO threshold power in resonators that are slightly over-coupled with coupling parameter $K = 3 \pm 1 = (1 + \sqrt{T})/(1 - \sqrt{T})$, where T is the transmission on-resonance [24].

We have also explored fine-scale frequency tuning and the frequency linewidth of the PhCR OPO; see Fig. 4. For frequency tuning, we leverage the thermo-optic coefficient of the tantala material so that a change in the chip temperature results in approximately -1 GHz/K shift of the PhCR modes [15]. By adjusting the temperature of the chip with a thermoelectric cooler and the frequency of the pump laser, we shift the frequency of the OPO output commensurate with the -1 GHz/K applied shift. Figure 4a presents the OPO spectra from two PhCR devices with different BG settings, hence they produce output at designed wavelengths of 1518 nm, 1595 nm, 1455 nm and 1672 nm ($\delta\nu_{OPO}$ = 9.53 THz and 26.74 THz) from a pump at 1555 nm. Specifically with the PhCR device that targets the latter wavelengths, we adjust the temperature in intervals of 10°C between 10°C and 60°C to affect output frequencies tuning with $dn/dT \approx -(n/\nu)\cdot(d\nu/dT) = 8.8 \times 10^{-6}$ $K^{-1}$, consistent with previous results [21].

We characterize the linewidth of the PhCR OPO tunable lasers, especially additive frequency noise that results from the nonlinear laser generation process itself. We perform the measurements with an unbalanced optical frequency discriminator Mach–Zehnder interferometer (MZI) setup of FSR 12.8 ± 0.3 MHz. In particular, we derive the integrated linewidth (Fig. 4b), using the $1/\pi$ phase and β-separation techniques [4,25,26] after measuring frequency noise. Following OPO laser generation, we optically filter the pump, OPO signal and idler and photodetect the individual wavelengths after propagating through the MZI and the output voltage from the photodetector is converted to frequency noise; see Fig. 4c. The pump laser has a $1/\pi$-integral linewidth of 13.7 ± 0.2 kHz and a β-separation linewidth of 693 ± 28 kHz, measured by our apparatus. The OPO signal and idler frequencies have a maximum $1/\pi$-integral linewidth of 34 ± 1 kHz and a β-separation linewidth of 958 ± 41 kHz; see Fig. 4b. We attribute the increase in OPO linewidth to thermorefractive noise (TRN) in the PhCR [27]. TRN jitter of the PhCR mode naturally induces pump detuning noise, which influences the OPO output frequency. Still, TRN couples to nonlinear processes in a relatively complex manner, especially for microcombs with large laser detuning in which the effect has been studied in detail [27]. Our present frequency noise measurements are in order of magnitude agreement with TRN models, accounting for the ~10 dB lower TRN due to tantala's smaller thermo-optic coefficient. In future experiments, we would explore the detuning dependence of TRN in PhCR OPOs, especially to understand the nonlinear coupling contribution [28,29].

We envision a range of potential applications for our OPO tunable-laser platform, especially in addressing requirements for wide laser-wavelength access. To understand both the operational system dynamics and the conversion efficiency of the pump laser power to OPO output power, we present analysis of output power as a function of the system parameters; see Fig. 5. Importantly, PhCR OPO operates bi-directionally, sending power in both the forward and backward directions with respect to the pump laser input. Experimentally, we detect OPO power bi-directionally by filtering the pump light as outlined in Fig. 2a. Figure 5a shows experimental OPO power in the forward and backward direction as a function of detuning relative to OPO threshold. We find that though the relative OPO powers are nearly equal in the forward (red) and backward (blue) directions, at certain detuning settings OPO power can propagate more predominantly in the forward or backward direction. Further, in the course of analyzing this data we observed that facet reflections play an important role in the forward and backward power balance.

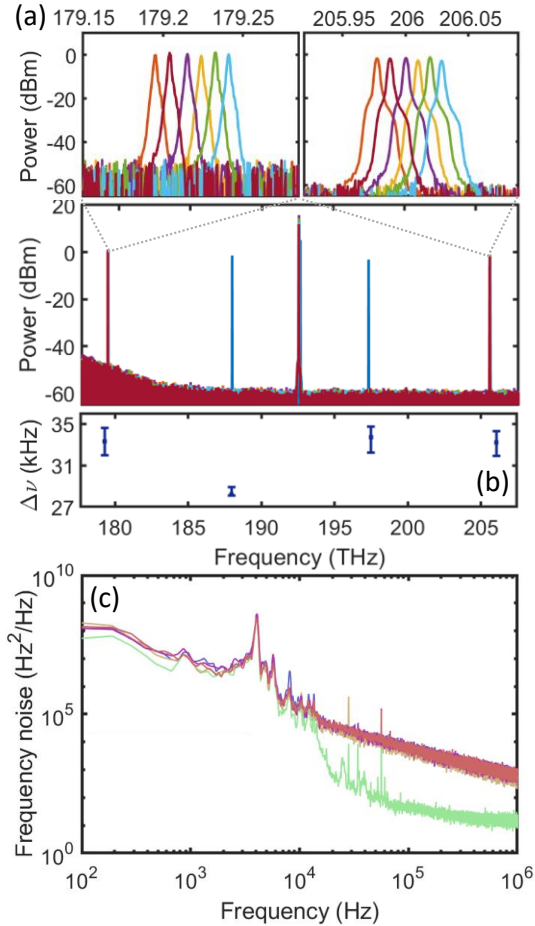

Fig. 4. (a) Optical spectra for two different BG with $\delta\nu_{OPO}$ = 9.53, 26.74 THz. Top left and right panels demonstrate thermo-optic tuning of OPO output as the PhCR is heated in 10 °C intervals from 10 °C to 60 °C while the pump laser is kept on resonance. (b) Measured OPO $1/\pi$-integrated linewidth, Δν, for idler and signal frequencies corresponding to (a). (c) Frequency noise measurements of OPO in (a,b). Green is the pump laser (193 THz), purple, gold, mauve and light red are OPO frequencies 198 THz, 188 THz, 206 THz and 179 THz, respectively.

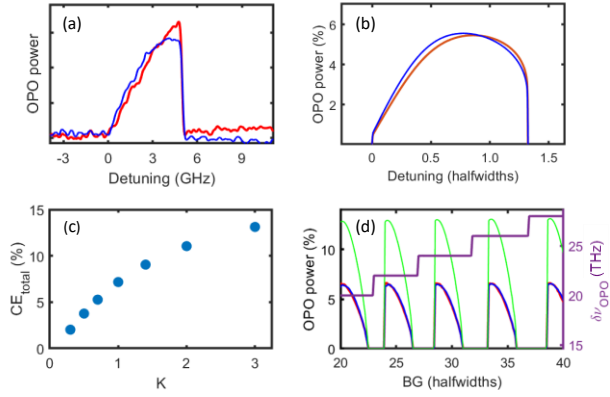

Fig. 5. (a) Experimental OPO power in the forward and backward direction (red and blue) for pump power ~ 1.3x OPO threshold as a function of laser detuning (relative to OPO threshold). (b) Theoretical OPO power for F = 2.28 in the forward and backward (red and blue) direction as a function of detuning. (c) Theoretical $CE_{total}$ as a function of coupling parameter, K, for F = 2.28. (d) OPO power dependence (left axis red, blue, and green OPO power in forward, backward, and total) and $\delta\nu_{OPO}$ (right, purple) as a function of BG.

To better understand the physics of PhCR OPO, we model the system with a pair of Lugiato-Lefever equations (LLE), $\partial_t \psi = -(1+i\alpha)\psi - \frac{i}{2}\beta \partial_\theta^2 \psi + i|\psi|^2\psi + F$, that describe the forward and backward intraresonator field $\psi(\theta, t)$, where t is time, and $\theta$ the angular coordinate along the resonator [30]. The parameters $\alpha$ and $\beta$ are the laser detuning and dispersion parameters, which are normalized to the PhCR mode linewidth and F is the pump field. The parameter $\beta$ includes the BG, and we incorporate fields propagating the both the forward and backward direction [31]. Our modelling also includes 6.8% reflection of optical power in either direction to reflect realistic experimental parameters, namely reflections from the chip facets. We model 2nd order dispersion of -0.094 halfwidths/mode and F = 2.28, which is distributed into both the forward and backward direction. In our model, the theoretical OPO threshold is near F = 2 in which the field splits between forward and backward direction and the overall dispersion nearly satisfies equation 2. However, the OPO threshold depends on the overall dispersion and BG setting ($\beta$) and the amount of field that is reflected towards the PhCR. Figure 5b presents theoretical OPO power predictions in the forward and backward direction as a function of laser detuning relative to OPO threshold and are in good qualitative agreement with experimental OPO power found in Fig. 5a. Note that the thermal bistability evident in Fig. 5a is not reflected in the theoretical model (Fig. 5b) [32]. Figure 5b predicts theoretical OPO power conversion ~5.5% into the forward and backward direction ($CE_{total}$ ~ 11%) when K = 3 (over-coupled), which agrees with the experimental results in Fig. 3b.

We also theoretically investigate the OPO conversion efficiency as a function of coupling parameter, K, and BG. We find that at BG setting with small phase-mismatch, $CE_{total}$ increases with K as seen in Fig 5c where we use a fixed field, F = 2.28 to reflect experimental parameters (($2.28/2)^2$ = 1.3). Note that though increasing K results in higher $CE_{total}$, this also increases the OPO threshold power [33]. Figure 5d presents PhCR OPO dynamics as a function of BG with fixed F = 2.28 and K = 3. The left axis shows the forward (red), backward (blue) and total (green) OPO conversion efficiency. At certain BG (e.g. 20 halfwidths in Fig. 5d), $CE_{total}$ peaks at 13% when the BG setting optimizes satisfies equation 2. Other values of BG (e.g. 37 halfwidths in Fig. 5d) provide poor phase-matching such that at F=2.28, no OPO is possible due to lack of sufficient parametric gain. The right axis in Fig. 5c shows $\delta\nu_{OPO}$ (purple), demonstrating BG dependent stepwise tuning of the laser output.

In conclusion, we have presented a platform for laser-wavelength access by use of PhCR OPO. We use a tantala integrated photonic platform to achieve by-design laser access >100 THz of optical bandwidth with narrow output linewidth. Our results demonstrate controllable laser wavelength generation with total pump-to-sideband conversion efficiencies > 10%, which is predicted by bi-directional LLE modelling.

**Funding.** This research has been funded by NIST, the DARPA LUMOS program, and AFOSR FA9550-20-1-0004 Project Number 19RT1019. This work is a contribution of the US Government and is not subject to copyright. Mention of specific companies or trade names is for scientific communication only and does not constitute a specific endorsement by NIST.

**Disclosures.** David Carlson is co-founder of Octave Photonics. The remaining authors do not currently have a financial interest in tantala integrated photonics.